\documentclass[twocolumn,showpacs,amsmath,amssymb,prb]{revtex4}

\usepackage{graphicx}
\usepackage[usenames]{color}

\newcommand{\Gr}{\ensuremath\Gamma_r}
\newcommand{\GL}{\ensuremath\Gamma_L}
\newcommand{\GR}{\ensuremath\Gamma_R}
\newcommand{\Gin}{\ensuremath\Gamma_\text{in}}
\newcommand{\Gout}{\ensuremath\Gamma_\text{out}}

\newcommand{\up}{\uparrow}
\newcommand{\dn}{\downarrow}

\begin{document}

\title{Spin-induced charge correlations in transport through interacting quantum dots with ferromagnetic leads}

\author{Stephan Lindebaum}
\affiliation{Theoretische Physik, Universit\"at Duisburg-Essen and CeNIDE, 47048 Duisburg}

\author{Daniel Urban}
\affiliation{Theoretische Physik, Universit\"at Duisburg-Essen and CeNIDE, 47048 Duisburg}

\author{J\"urgen K\"onig}
\affiliation{Theoretische Physik, Universit\"at Duisburg-Essen and CeNIDE, 47048 Duisburg}

\date{\today}
%72.70.+m	Noise Processes and Phenomena
%73.23.Hk	Coulomb Blockade; Single Electron Tunneling
%85.35.Gv	Single electron devices
%85.75.-d 	Magnetoelectronics; spintronics: devices exploiting spin polarized transport or integrated magnetic fields

\begin{abstract}
We study the full counting statistics of electronic transport through a single-level quantum dot weakly coupled to two leads, with either one or both of them being ferromagnetic.
The interplay of Coulomb interaction and finite spin polarization implies spin-correlation induced charge correlations that give rise to super-Poissonian transport behavior and positive cross correlations of the currents of the two spin species.
In the case of two ferromagnetic leads, we analyze the nontrivial dependence of the cumulants on the angle between the polarization directions of the leads.
We find diverging second and higher cumulants for spin polarizations approaching unity.
\end{abstract}

\maketitle

%%SECTION INTRODUCTION
\section{Introduction}
The possibility to control not only charge but also spin currents defines an important goal in the field of spintronics.\cite{datta:1990,zutic:2004}
Predicted effects, such as the tunnel magnetoresistance (TMR),\cite{julliere:1975} have already proven industrial relevance.
Transport through mesoscopic systems is, on the other hand, strongly influenced by Coulomb interaction effects.
It is, therefore, an important issue to understand the interplay between Coulomb interaction and finite spin polarization in nanoscale devices.
	
Quantum dots (QDs) attached to ferromagnetic leads are a convenient model system to study the implications of this interplay.
This includes quantum-dot spin valves, in which transport through a quantum dot depends on the relative orientation of the magnetization directions of the source and the drain leads.
Recent theoretical works report on a variety of complex transport properties and effects, such as
negative differential conductance,\cite{bulka:2000,braun:2004,rudzinski:2005,braig:2005}
spin precession,\cite{koenig:2003,braun:2004,rudzinski:2005,braig:2005}
inverse TMR effect,\cite{rudzinski:2005,inverseTMR1,inverseTMR2,inverseTMR3}
shot noise,\cite{weymann:TMR,souza:2008}
spin-diode behavior,\cite{weymann:spindiode}
and the existence of an interaction-induced exchange field between leads and QD, which leads to a precession of the accumulated dot spin\cite{koenig:2003,braun:2004} or a splitting in the Kondo 
resonance.\cite{martinek:2003}
The latter has been experimentally confirmed recently.\cite{pasupathy:2004,hamaya,hauptmann} 
Further experimental studies include spin-dependent transport through metallic nanoclusters,\cite{schelp:1997,yakushiji:2001,deshmukh:2002,zhang:2005,bernand:2006} and quantum dots realized in carbon nanotubes.\cite{sahoo,hauptmann,Merchant:2008} 

	Transport through mesoscopic devices is of stochastic nature, i.e., the charge current fluctuates. A full description of transport is, thus, only given by the knowledge of the probability distribution $P(N ,t_{0})$ that $N$ electrons have passed through the system in time $t_{0}$. The full counting statistics (FCS) is obtained from the cumulant generating function (CGF) $S(\chi)$ that is related to the probability distribution by
\begin{equation}
		S(\chi) = \ln \left[ \sum_{N=-\infty}^{\infty} e^{iN \chi} P(N,t_{0}) \right] \, .
\end{equation}
From the CGF the cumulants of the current can be obtained by performing derivatives with respect to the counting field, $\left. \kappa^{(n)}=(-i)^{n}(e^{n}/t_{0})\partial^{n}_{\chi} S(\chi) \right|_{\chi=0}$. In the long-time limit, the first four cumulants are related to the average current, the (zero-frequency) current noise, the skewness, and the kurtosis.
If transport is carried by uncorrelated and rare processes, then the statistics will be Poissonian and the cumulants normalized with respect to average current and elementary charge, $\kappa^{(n)}/(e^{n-1}\kappa^{(1)})$, will be one (for $n=2$ this ratio defines the Fano factor $F$).
Correlations may lead to smaller or larger values, i.e., sub- or super-Poissonian statistics.

Recent progress in the theoretical description of FCS in electronic transport through nanostructures has been achieved by including interaction effects,\cite{interactingFCS1,interactingFCS2,interactingFCS3,interactingFCS4,bagrets:2003}
interference effects,\cite{interferenceFCS1,interferenceFCS2,interferenceFCS3}
frequency-dependent FCS,\cite{frequenceFCS}
and the description of non-Markovian effects.\cite{braggio:2006}
The measurement of FCS has also become possible in quantum dots through real-time detection of electrons by means of quantum point contacts. \cite{FCSExpQPC1,FCSExpQPC2,FCSExpQPC3}
	
\begin{figure}
	\includegraphics[width=.7\columnwidth]{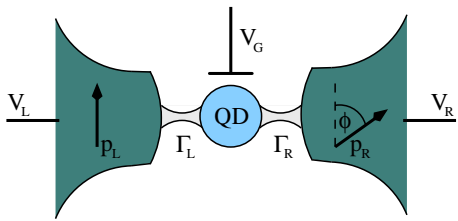}
	\caption{(Color online) A quantum-dot spin valve consists of a QD connected to two ferromagnetic leads (F-D-F), whose polarizations ${\bf p}_{r}$ enclose a tunable angle $\phi$. If the polarization of one of the leads vanishes, the structure defines a F-D-N system.}
	\label{fig:system}
\end{figure}

In the following, we consider transport through a single-level quantum dot in the limit of weak tunnel coupling between dot and leads (Fig.~\ref{fig:system}). Furthermore, we are interested in the shot-noise regime and the long-time limit.
We, therefore, assume a large bias voltage such that only transport processes in one direction are relevant. 
For nonmagnetic leads, i.e., a N-D-N system, and in the absence of Coulomb interaction, the two spin channels are independent of each other. Each of the two channels is described by the CGF (Ref. \onlinecite{bagrets:2003})
\begin{eqnarray}\label{eq:BN}
	S(\chi)&=&-t_0\, \alpha \left[ 1-\sqrt{ 1+ \beta (e^{i\chi}-1) } \right] \, ,
\end{eqnarray}
with positive coefficients $\alpha = (\Gin + \Gout)/2$ and $\beta=4\Gin\Gout/(\Gin+\Gout)^2$, where $\Gin/\hbar$ and $\Gout/\hbar$ are the rates for an electron tunneling in and out of the dot, respectively.
The total CGF for charge transport is just the sum of the two equal spin contributions, which simply introduces an overall factor of $2$.
It always describes sub-Poissonian transport.
For example, the Fano factor, $F= 1 - \beta/2$ is always limited by the Poissonian value of one.
The latter is approached for $\beta \rightarrow 0$, which is achieved for very asymmetric tunnel couplings to the left and right leads, a situation in which the system behaves like a single barrier.

The presence of interaction in the N-D-N system with a spin-degenerate level in the quantum dot is not yet sufficient to generate correlations with super-Poissonian statistics.
In contrast, it turns out that for $U=\infty$ the CGF still has the structure of Eq.~({\ref{eq:BN}) with $\Gin = 2\GL$ and $\Gout=\GR$, where only transport from the left ($L$) to the right ($R$) lead, with rates $\GL/\hbar$ and $\GR/\hbar$, respectively, is possible.
The factor of $2$ in $\Gin$ accounts for the fact that there are two possible spins that can enter an empty dot while for a singly-occupied dot there is only one possible spin to tunnel out.
To achieve super-Poissonian statistics, one may, in addition, break the spin symmetry by introducing a finite Zeeman term.\cite{sukhorukov,belzig:2005}
In the present paper, we will follow an alternative route, namely, to make use of one or two ferromagnetic leads, similar to Refs.~\onlinecite{cottetRevB} and \onlinecite{cottetRevLett}, where three-terminal devices involving ferromagnetic leads were considered.
The Coulomb interaction will correlate the different spin channels that will, in some circumstances, induce charge correlations.
It is, thus, the combination of the two ingredients, finite spin polarization and Coulomb interaction, which will lead to the appearance of super-Poissonian statistics.

The specific system we consider in this paper is a single-level QD weakly coupled to one or two ferromagnetic leads (see Fig.~\ref{fig:system}). 
To analyze the role of Coulomb interaction, we compare the two limits of either noninteracting electrons or strong Coulomb interaction such that double occupancy of the dot is prohibited.
In the case of two ferromagnetic leads, the interplay of Coulomb interaction and finite spin polarization gives rise to a nontrivial dependence of the cumulants on the angle between the magnetization directions of the leads.
In particular, we find strongly super-Poissonian behavior with diverging higher cumulants for large spin polarization in the leads.
But already the system with one ferromagnetic and one normal lead coupled to a quantum dot (F-D-N) displays super-Poissonian behavior if the electrons are injected from the normal electrode.
Although for transport in the opposite direction, the statistics remains sub-Poissonian.

%%SECTION SYSTEM
\section{System and Method}	\label{sec:system}
	The quantum-dot spin valve shown in Fig.~\ref{fig:system} is modeled by the Hamiltonian
	\begin{equation}\label{eq:hamiltonian}
		H = H_\text{dot} + H_L+H_R +H_T\, .
	\end{equation}
	The first part, $H_\text{dot} = \sum_\sigma \varepsilon\, c_\sigma^\dagger c_\sigma + U n_{\uparrow}n_{\downarrow}$, describes the QD as an Anderson impurity with a spin-degenerate electronic level $\varepsilon$ and charging energy $U$ for double occupation. Each of the leads is described as a reservoir of noninteracting fermions $ H_r = \sum_{k{s}} \varepsilon_{rk{s}}^{}\, a_{rk{s}}^\dagger a_{rk{s}} $ with indices for lead $r \in \{L,R\}$ and momentum $k$. The index ${s}=+(-)$ denotes the majority (minority) spin states with the density of states $\rho_{r}^{{s}}$. The lead polarization is characterized by the direction of the polarization vector ${\bf p}_{r}$ and its magnitude $p_r=|{\bf p}_{r}| = (\rho_{r}^{+}-\rho_{r}^{-})/(\rho_{r}^{+}+\rho_{r}^{-})$. The tunneling Hamiltonian $H_T = \sum_r H_{T,r}$\;, with
	\begin{equation}
		H_{T,r}	=	\sum_{k{s}\sigma}     \,V^{r}_{k{s}\sigma}\,       a^\dag_{rk{s}} c_{\sigma}^{} + \text{H.c.}\; ,
	\end{equation}
describes tunneling between dot and lead $r$.
	Due to the fact that the quantization axes in the leads cannot, in general, both coincide with that of the dot, the tunnel matrix elements $V^{r}_{ks\sigma}$ consist not only of the (spin-independent) tunnel amplitude $t_r$ but also contain an SU(2) rotation about the relative polar angles $\theta_{r}$ and $\phi_{r}$ between lead polarization and dot quantization axis. Choosing the quantization axis ${\bf \hat{e}}_{z}=({\bf p}_{{L}}\times{\bf p}_{R})/|{\bf p}_{L}\times{\bf p}_{R}|$ of the QD spin orthogonal to both lead polarization directions, the tunneling Hamiltonian for the left lead becomes
	\begin{eqnarray}\label{eq:HT}
		H_{T,L}	&=&\frac{t_L}{\sqrt{2}}\sum_{k}	a^{\dag}_{Lk+} \left( e^{i\phi/4}c_{\up} + e^{-i\phi/4} c_{\dn} \right)\nonumber\\
		& &	+a^{\dag}_{Lk-} \left( -e^{i\phi/4}c_{\up} + e^{-i\phi/4} c_{\dn} \right)  + \text{H.c.}\;,
	\end{eqnarray}
	while the right lead is described by the same expression but with the replacements $L\rightarrow R$ and $\phi\rightarrow -\phi$. The tunneling rate for electrons with spin $s=\pm$ is quantified by $ \Gamma_r^{\pm}/\hbar =  2 \pi \left|t_{r}\right|^2 \rho_{r}^{\pm}/\hbar = \Gamma_{r}(1\pm p_{r})/\hbar $. For simplicity, we assume the density of states~$\rho^s_r$ and the tunneling amplitudes~$t_r$ to be independent of wave vector and energy.
Furthermore, we define $\Gamma_r \equiv (\Gamma_{r}^{+}+\Gamma_{r}^{-})/2$ as well as $\Gamma\equiv \sum_r \Gamma_r$.

	The reduced density matrix for the QD degrees of freedom, $\langle |\nu \rangle \langle \mu | \rangle$, where $\mu$ and $\nu$ label the QD states, contains, in general, both diagonal and off-diagonal matrix elements. The diagonal components $P_{0}$, $P_{1}$, and $P_{d}$ describe the probabilities to find the dot empty, singly, or doubly occupied, respectively. The average spin on the QD, with components $S_{x}$, $S_{y}$, and $S_{z}$, contains off-diagonal density-matrix elements as well. 
We summarize these six quantities, containing both diagonal and off-diagonal density-matrix elements,\cite{interferenceFCS3} in a vector ${\bf p}=(P_{0},P_{1},P_{d},S_{x},S_{y},S_{z})$. 
Its time evolution is described by an $N$-resolved kinetic equation
	\begin{equation}\label{eq:meq}
		\frac{d}{dt}{\bf p}(N,t)  =   \sum_{N'} \int_0^{t} dt' \;{\bf W}(N-N',t-t')\; {\bf p}(N',t')\;,
	\end{equation}
where $N$ is the number of transferred electrons.

	The kernel ${\bf W}(N-N',t-t')$ of the kinetic equation can be obtained using a diagrammatic real-time technique formulated on the Keldysh contour. It allows for a systematic perturbative expansion in the coupling strength. In this paper, we truncate the expansion at the lowest order $\Gamma$ to describe the weak-coupling limit (sequential tunneling). For a detailed derivation of this diagrammatic language and its rules for the calculation of diagrams, we refer to Refs.~\onlinecite{diagrams1,diagrams2,technique1,technique2}. Not described in these references is the inclusion of the counting field~$\chi$, the Fourier-conjugated variable of the number of transferred electrons $N$. It is introduced by multiplying the tunnel amplitudes with phase factors $e^{\pm i\chi/2}$, where the sign is chosen such that $t_{L} \rightarrow t_{L} e^{i\chi/2}$ and $t_{R} \rightarrow t_{R} e^{-i\chi/2}$ if the tunnel vertex is placed on the upper, and $t_{L} \rightarrow t_{L} e^{-i\chi/2}$ and $t_{R} \rightarrow t_{R} e^{i\chi/2}$ if it is on the lower branch of the Keldysh contour.
		
		As has been described in Ref.~\onlinecite{braggio:2006} the solution of the kinetic equation [Eq.~(\ref{eq:meq})] can be found by Fourier transformation with respect to $N$ (which introduces the counting field $\chi$) and Laplace transformation with respect to time $t$ (which introduces the variable~$z$). From the solution ${\bf p}(\chi,z)$ one obtains the CGF~$S(\chi)$. In general, the CGF includes non-Markovian corrections, related to a finite support of the kernel $\mathbf{W}(N-N',t-t')$ in time, i.e., a $z$ dependence of the Fourier transform. However, it has been shown\cite{braggio:2006} that these non-Markovian corrections do not enter the CGF for the lowest-order term of a perturbation expansion in some small parameter, which in our case is provided by the coupling strength $\Gamma$. 
In that case, the CGF is given as the $z=0^+$ limit of the eigenvalue $\lambda(\chi,z)$ of the kernel $\bf W$, whose real part has the smallest absolute value,
	\begin{eqnarray}\label{eq:Smarkov}
		S(\chi)=t_{0} \lambda(\chi)\;,
	\end{eqnarray}
which was first derived in Ref.~\onlinecite{bagrets:2003}.

	The kernels of the systems considered in the present paper are shown in the Appendix. In some cases, it is possible to find simple analytic expressions for the full $\chi$ dependence of the sought-for eigenvalue $\lambda(\chi)$, from which one can calculate all cumulants. For cases, in which such an analytic solution of the eigenvalue problem is not accessible,  Flindt {\em et al.}\cite{flindt:2008} suggested an alternative route, which is based on a Rayleigh-Schr\"odinger perturbation theory to expand the eigenvalue in $\chi$ and, thus, allows for a calculation of the cumulants without the need to solve the full eigenvalue problem. We compute the full analytic solution of the CGF in Secs. \ref{sec:FDN_U=0}, \ref{sec:FDN}, and \ref{sec:QDSV_U=0}, while in \ref{sec:NDF} and \ref{sec:QDSVU}, we only use the  perturbative solution.
	
%%SECTION FDN		
\section{F-D-N system}	\label{sec:oneFM}
	We start by considering a \mbox{F-D-N} system, i.e., only one of the leads is ferromagnetic with spin polarization~$p$. To set a reference, we first consider the case of noninteracting electrons, $U=0$. Afterward, we study the limit of strong Coulomb interaction, $U=\infty$. For the latter limit, the direction of electron transfer, from ferromagnet to normal lead or vice versa, will have an import influence on the FCS. In both limits we constrain ourselves to the shot-noise regime, in which only unidirectional transport is possible. Hence, the Fermi functions are $f_r(\varepsilon)=1$ and $0$ if $r$ refers to the source and drain leads, respectively. Furthermore, in the limit of strong Coulomb interaction, we put $f_r(\varepsilon+U)=0$.

In order to analyze how the two spin channels contribute, we introduce a spin-resolved CGF with different counting fields $\chi_\sigma$ for the two spin species $\sigma=\up,\dn$. We define $\up$ to be the majority spin of the ferromagnet. In this basis, no coherent superpositions between up and down spins occur for the considered system (unlike the quantum-dot spin valve studied in Sec.\ref{sec:QDSV}), and the generalization to spin-resolved counting is straightforward. 
In addition to the cumulants we investigate the zero-frequency cross correlations of the two spin currents
$C_{\up\dn} = \int dt \left[\left<I_\up(t) I_\dn(0)\right>-\left<I_\up(t)\right>\left<I_\dn(0)\right>\right]$.
They can be obtained from the spin-resolved generating function by differentiation, $C_{\up\dn} = -(e^2/t_0) \left.\partial_{\chi_{\up}}\partial_{\chi_{\dn}} S(\chi_{\up},\chi_{\dn})\right|_{\chi_\up=\chi_{\dn}=0}$.
The idea of spin-resolved counting has already been employed in the context of detection of spin singlets,\cite{dilorenzo} and transport between superconductors and ferromagnets.\cite{spinresolved1,spinresolved2}

	\subsection{Noninteracting dot -- $U=0$}\label{sec:FDN_U=0}
In the absence of Coulomb interaction on the QD, both spin channels contribute to transport independently and the CGF is just the sum of the individual CGFs.
These individual CGFs are given by Eq.~(\ref{eq:BN}) with tunneling rates $\Gamma_{\text{in},\text{out}}=\Gamma_{N},\Gamma_{F}^{\sigma}$, which implies $\alpha=(\Gamma_{F}^{\sigma}+\Gamma_N)/2$, and $\beta=4\Gamma_{F}^{\sigma}\Gamma_N/(\Gamma_{F}^{\sigma}+\Gamma_N)^2$, independent of the direction of the applied bias voltage. Here, $\Gamma_{F}^{\sigma}=\Gamma_F(1\pm p)$ is the coupling strength of the ferromagnet to the dot for the majority-spin ($+$ sign) and minority-spin electrons ($-$ sign), and $\Gamma_N$ is the coupling strength to the normal lead.

	In Fig.~\ref{fig:FNDU0}, we show the first three (normalized) cumulants as a function of the degree of spin polarization $p$. In addition to the cumulants for the charge transport, we display separately the cumulants for the two spin channels, normalized with respect to the current of the respective channel. We find that the FCS is always sub-Poissonian. For strong polarizations $p$ the FCS is mainly determined by the majority-spin carriers, in the sense that the dependence of the total cumulants on $p$ resembles that of the majority-spin channel, and both curves coincide for $p\rightarrow1$.
	Since the two spin species are transferred independently, their cross correlation is found to vanish, $C_{\up\dn,U=0}=0$.
	
		\begin{figure}
			\includegraphics[width=.95\columnwidth]{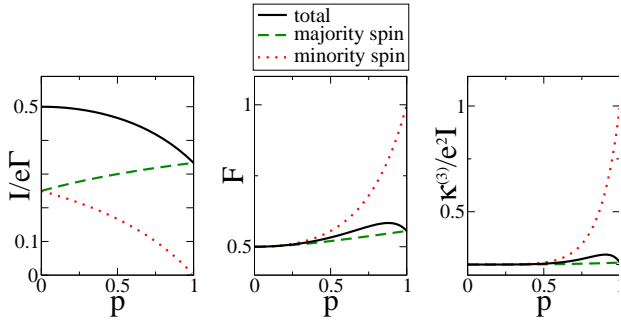}
			\caption{(Color online) Cumulants of a symmetric ($\Gamma_F=\Gamma_N=\Gamma/2$) noninteracting F-D-N system.}
		\label{fig:FNDU0}
		\end{figure}

	\subsection{Strong Coulomb interaction -- $U=\infty$}
		In the presence of strong Coulomb interaction on the dot, the spin channels can no longer be regarded as independent. Correlated transport processes of the two spin states take place. It, then, makes a difference whether the ferromagnetic lead serves as source or drain. As it turns out, super-Poissonian behavior is obtained only when the electrons are injected from the normal electrode. 

		\subsubsection{Injection from the ferromagnet}\label{sec:FDN}
			If electrons are injected into the QD from the ferromagnet, the FCS can still be described by Eq.~(\ref{eq:BN}) with $\Gin= 2\Gamma_F = \Gamma_{F}^{\up}+\Gamma_{F}^{\dn}$ and $\Gout=\Gamma_N$ as for the \mbox{N-D-N} system.
In fact, the degree of spin polarization $p$ does not enter at all the CGF, and the statistics remains sub-Poissonian.
The underlying reason for this is that all tunneling rates are affected exclusively by either the finite spin polarization of the lead or the Coulomb interaction on the dot but never by both of them. 
For the transition between an empty and a singly-occupied dot the spin polarization of the ferromagnet matters but the charging energy for double occupation does not.
On the other hand, when the dot is singly occupied, the strong Coulomb interaction allows only the dot electron to tunnel out into the normal lead for which the tunneling rates are spin independent.

The spin-resolved analysis reveals that the larger contribution to the total statistics comes from majority spins, just as in a noninteracting system.
For $p \rightarrow 1$ minority spins are increasingly rarely injected from the source so that their statistics becomes Poissonian. It is known\cite{buttiker} that the zero-frequency current cross correlations are always negative for noninteracting electronic circuits with leads in thermal equilibrium and constant applied bias voltage. Here, we obtain negative cross correlations also in a strongly interacting system (Fig.\ref{fig:FDNCC}):
				\begin{equation}
					C^\text{FDN}_{\up\dn,U=\infty}=-\frac{2e^2(1-p^2)\Gamma_F^2\Gamma_N^2}{(\Gamma_F+\Gamma_N)^3}.
				\end{equation}
				\begin{figure}
					\includegraphics[width=1\columnwidth]{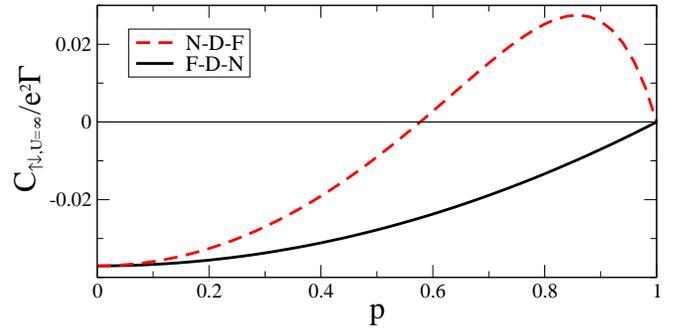}
					\caption{(Color online) Cross correlations between the current fluctuations of the two spin species in the strongly interacting system with one ferromagnetic lead for both transport directions ($\Gamma_F=\Gamma_N=\Gamma/2$).}
				\label{fig:FDNCC}
				\end{figure}

		\subsubsection{Injection from the normal lead}\label{sec:NDF}
For reversed transport voltages, where spins are injected from the normal metal, the degree of polarization of the ferromagnet plays an important role. Then, tunneling into the dot is spin independent, but the drain contact is spin sensitive. A minority spin occupying the dot leads to an interruption of the electron stream (spin blockade), which causes electron bunching and, thus, enhanced noise, see Fig.~\ref{fig:NDFUinf}. The Fano factor is dominated by the majority electrons and rises to three for perfect polarization.

			This Fano factor can be understood by the following argument. The probability that a majority spin enters the empty dot is $1/2$. Since they have a short dwell time, several majority electrons are transported in quick succession until a minority spin enters the dot (also with probability $1/2$) and blocks transport. Therefore, the probability that $N$ electrons are transferred during such a process is $1/2^N$. These characteristics result in a Fano factor of~3.\cite{belzig:2005,Fano31}
			\begin{figure}
				\includegraphics[width=.95\columnwidth]{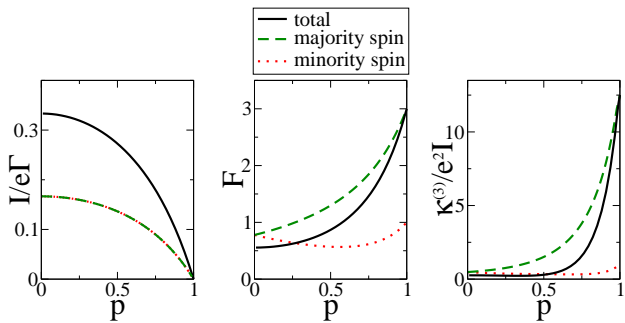}
				\caption{(Color online) Cumulants of the \mbox{N-D-F}-system in which the electrons are injected from the normal electrode. The total current [solid (black)] splits up into two equal spin currents [dashed (green) and dotted (red)] and the noise is enhanced due to bunching. For higher cumulants the enhancement increases ($\Gamma_F=\Gamma_N=\Gamma/2$).}
			\label{fig:NDFUinf}
			\end{figure}
			
			We obtain the following analytic expressions for the average current and the Fano factor:
			\begin{eqnarray}
				I_{U=\infty}^\text{\begin{tiny}NDF\end{tiny}}&=&\frac{2e(1-p^2)\Gamma_F\Gamma_N}{2\Gamma_N+(1-p^2)\Gamma_F}\,,\\
				F_{U=\infty}^\text{\begin{tiny}NDF\end{tiny}}&=&\frac{4(1+2p^2)\Gamma_N^2+(1-p^2)^2\Gamma_F^2}{[2\Gamma_N+(1-p^2)\Gamma_F]^{2}}\,.
			\end{eqnarray}

			Again, the spin-resolved analysis reveals that the majority spins govern the total statistics and the statistics of minority spins turns Poissonian for $p\rightarrow 1$, see Fig.~\ref{fig:NDFUinf}.
However, because of interaction and bunching there is no interpretation of the origin of this effect as obvious as in the two systems discussed in the previous paragraphs.
			We obtain the following expression for the cross correlations of the two spin species,				\begin{equation}
					C^\text{NDF}_{\up\dn,U=\infty}=-\frac{2e^2(1-p^2)\left[(1-p^2)\Gamma_N-2p^2\Gamma_F\right]\Gamma_N\Gamma_F^2}{\left[(1-p^2)\Gamma_N+2\Gamma_F\right]^3},
				\end{equation}
			which is plotted in the limit of symmetric tunneling rates in Fig.~\ref{fig:FDNCC}. We find  positive cross correlations for polarizations larger than $1/\sqrt{3}$, which in our model coincides with the regime of super-Poissonian Fano factors. This result is similar to that in three-terminal quantum-dot devices, for which positive cross correlations have been found as a consequence of dynamical spin blockade.\cite{cottetRevLett,cottetRevB}

%%SECTION SPIN VALVE
\section{Quantum-Dot Spin Valve (F-D-F)}    \label{sec:QDSV}	
	A quantum-dot spin valve consists of two ferromagnetic leads coupled to a QD. For simplicity, we constrain ourselves in the following to symmetric polarizations ($p_{L}=p_{R}=p$) and the shot-noise regime in which only transport from the left to the right lead is possible. The Fermi functions are, therefore, $f_L(\varepsilon)=1$ and $f_R(\varepsilon)=0$, as well as $f_L(\varepsilon+U)=f_R(\varepsilon+U)=0$ in the limit of strong Coulomb interaction.
The electric current through the dot depends on the relative angle $\phi$ between the magnetization directions of the leads: it is maximal for parallel and minimal for antiparallel alignment.
	In this section, contrary to the F-D-N system, we do not study the spin-resolved FCS to avoid destroying the coherences between spin-up and spin-down electrons.

	\subsection{Noninteracting dot -- $U=0$}\label{sec:QDSV_U=0}
	We first consider the limit of noninteracting electrons to create a reference for the strongly interacting situation. Although Coulomb interaction is absent, we cannot, in general, separate the transport into two independent spin channels. The reason is the noncollinearity of the leads' magnetization directions, which yields that for any choice of the spin-quantization axis the two spin channels are coupled to each other. Therefore, the CGF acquires, in general, a more complicated form different from Eq.~(\ref{eq:BN}). Still the transport remains sub-Poissonian. In the special limit of symmetric tunnel couplings, $\GL=\GR=\Gamma/2$, the CGF simplifies to the form of Eq.~(\ref{eq:BN}), with $\alpha=\Gamma$, and $\beta=1-p^2(1-\cos\phi)/2$. 
		
		The angle dependence of the first four (normalized) cumulants is shown in Fig.~\ref{fig:QDSVU0} for three different values of the lead polarization $p$. For $\phi\neq0$ the transparency is reduced due to the spin-valve effect. This results in enhanced second and higher cumulants. They approach the Poissonian limit for $\phi=\pi$ and $p=1$, for which the transmission goes to zero. We remark that in the limit $p\rightarrow 1$ the model can be mapped onto a double-dot geometry with spinless noninteracting electrons with identical full counting statistics.\cite{interferenceFCS3}

		\begin{figure}
			\includegraphics[width=.95\columnwidth]{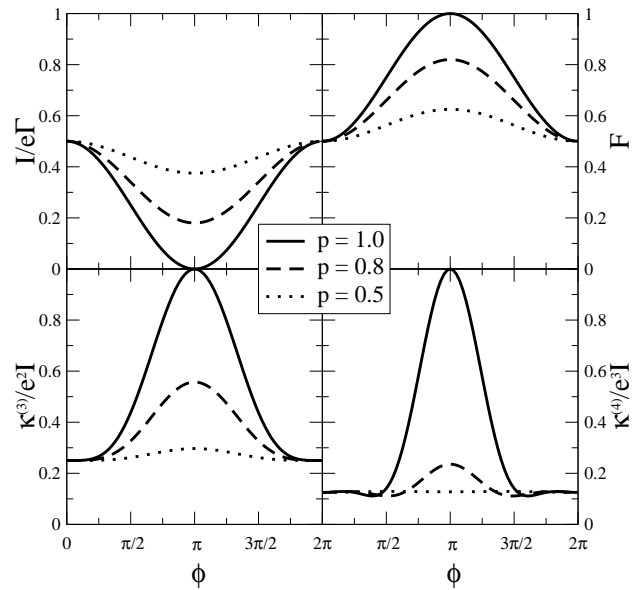}
			\caption{Cumulants of the symmetric ($p_{L}=p_{R}=p$, $\GL=\GR=\Gamma/2$), noninteracting quantum-dot spin valve as a function of the angle $\phi$. For \mbox{$\phi=0$} the statistics of noninteracting particles in a double barrier system is assumed, while for \mbox{$\phi=\pi$} transport is Poissonian due to spin blockade.}
			\label{fig:QDSVU0}
		\end{figure}

	\subsection{Strong Coulomb interaction -- $U=\infty$}\label{sec:QDSVU}
	
	The situation is qualitatively different when strong Coulomb interaction is taken into account. Since the analytic formulas are rather complicated, we only present the numerical results of the $\phi$ dependence of the first four (normalized) cumulants in Fig.~\ref{fig:QDSVinfarbang}. We find a nontrivial angular dependence. In particular, we obtain super-Poissonian transport behavior for a large parameter range. Furthermore, we see that the second and higher cumulants even diverge for high polarizations $p\rightarrow 1$ and small angles $\phi\rightarrow 0$. This dramatic effect is a consequence of the interplay of spin polarization and Coulomb interaction. The divergence (enhancement) increases for higher cumulants.
		\begin{figure}
			\includegraphics[width=.95\columnwidth]{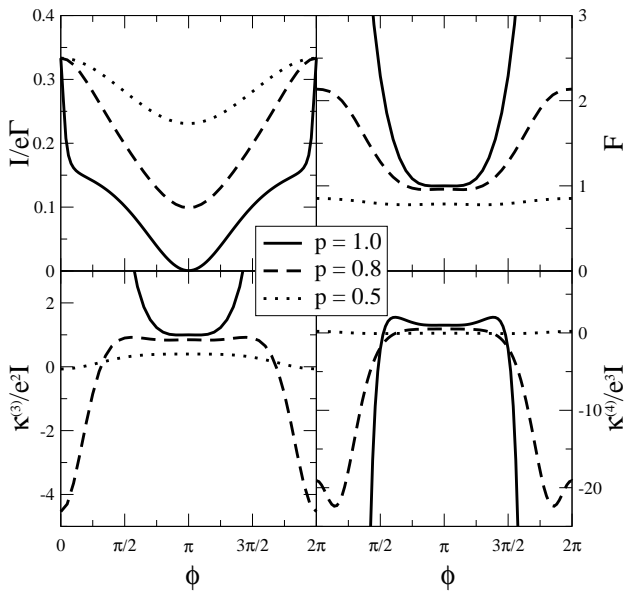}
			\caption{Cumulants of the strongly interacting quantum-dot spin valve with symmetric parameters: $\GL=\GR=\Gamma/2$, $p_{L}=p_{R}=p$. In the parallel case ($\phi=0$) Coulomb blockade of minority spins leads to bunching. In the antiparallel situation (\mbox{$\phi=\pi$}) the statistics becomes Poissonian for $p\rightarrow1$ due to spin blockade.}
		\label{fig:QDSVinfarbang}
		\end{figure}

		To understand the underlying mechanism for this effect, let us consider the case of parallel magnetizations, $\phi=0$. Due to the infinite charging energy, the current is reduced by a factor $(\GL+\GR)/(2\GL+\GR)$ as compared to the noninteracting case.\cite{bagrets:2003}
Switching on a finite spin polarization $p$ does not change the value of the current:
on one hand, the rates for the majority electrons to enter, $\GL(1+p)$, or to leave, $\GR(1+p)$, is increased but at the same time, the rates for the minority electrons are reduced, $\GL(1-p)$ and $\GR(1-p)$. Once the latter enter the dot, they remain for a long time, thus blocking transport for the majority-spin channel. This means that the majority electrons are bunched, and an elevated noise is expected. (This is reminiscent of the effects described in Refs.~\onlinecite{djuric:2005} and \onlinecite{li:2008}, where double-dot systems were considered.) In fact, the analytic expressions
\begin{eqnarray}
			I_{U=\infty,\phi=0}&=&\frac{2e\GL\GR}{2\GL+\GR},\\
			F_{U=\infty,\phi=0}&=&\frac{4\frac{1+p^2}{1-p^2}\GL^2+\GR^2}{(2\GL+\GR)^2} \, ,
		\end{eqnarray}
for the current and the Fano factor\cite{braun:2006} yield a divergence of $F$ as $p\rightarrow 1$.
		
		For the antiparallel arrangement bunching is not relevant since both spin species now experience equal coupling strengths (with only the roles of source and drain exchanged). For finite polarizations $p$, the noise is enhanced as compared to the case $p=0$. The reason for this is a suppression of transport due to spin blockade: the majority spin of the source lead is the minority spin of the drain and, thus, can hardly leave the dot to the drain. This leads to a reduction in the average transported charge, while the statistics is determined by the bottleneck of electrons leaving the dot. The noise remains sub-Poissonian, reaching the Poissonian limit for $p\rightarrow 1$. The analytic  expressions for the current and the Fano factor are
		\begin{eqnarray}
			I_{U=\infty,\phi=\pi}&=&\frac{2e\GL\GR}{2\frac{1+p^2}{1-p^2}\GL+\GR}\;,	\\
			F_{U=\infty,\phi=\pi}&=&\frac{4\frac{1+4p^2-p^4}{(1-p^2)^2}\GL^2+\GR^2}{(2\frac{1+p^2}{1-p^2}\GL+\GR)^2} \, .
		\end{eqnarray}

		It has been pointed out that the tunnel coupling of the QD levels to spin-polarized leads induces an effective exchange field experienced by the QD spins.\cite{koenig:2003,braun:2004,martinek:2003} 
This exchange field gives rise to a precession of an accumulated QD spin that, in turn, modifies the transport characteristics. To investigate the influence of this exchange field we compare our results to the case when the exchange field is set to zero by hand. This comparison is shown in Fig.~\ref{fig:QDSVinfB}.
Apart from the points $\phi=0$ and $\phi=\pi$ (where the exchange field is collinear to the accumulated spin and, therefore, has no impact), the precession of the QD spin tends to lift any spin blockade, which increases the current and also reduces electron bunching, so that the Fano factor and higher normalized cumulants are decreased.
		\begin{figure}
			\includegraphics[width=.95\columnwidth]{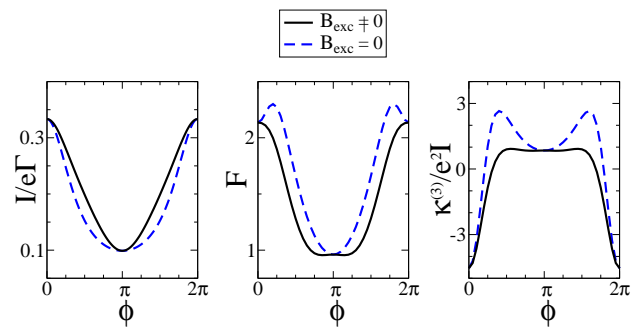}
			\caption{(Color online) The exchange field increases the current of a strongly interacting quantum-dot spin valve and smoothens sharp characteristics of the cumulants. Parameters: $p_{L}=p_{R}=0.8$, $\GL=\GR=\Gamma/2$, $\mu_{L/R}=\pm15k_\text{B}T$, and $\varepsilon=0$.}
		\label{fig:QDSVinfB}
		\end{figure}

%%SECTION CONCLUSION
\section{Conclusion}    \label{sec:conclusion}

	We have investigated the FCS of electronic transport through a QD coupled to ferromagnetic and normal leads by means of a diagrammatic real-time technique. As a main result we found that the interplay of finite spin polarization in the electrodes and strong Coulomb interaction on the QD can lead to super-Poissonian transport statistics, with higher cumulants being more and more enhanced. For the system of a QD coupled to one normal and one ferromagnetic lead, super-Poissonian behavior only appeared for transport from the normal lead to the ferromagnet, associated with positive cross correlations of the spin channels.
The most dramatic effect was expected for the quantum-dot spin valve, in which both leads were ferromagnetic.
In this case, the second and higher cumulants diverged for small angles between the leads' magnetization directions and large polarizations. This effect was understood to originate from bunching.

%%SECTION ACKNOWLEDGEMENTS	
\acknowledgments
	We acknowledge financial support from DFG via Contracts No. SFB 491, No. SPP 1285, and No. KO 1987/4.

\appendix

%%SECTION KERNEL
\section{Kernels of kinetic equations}
	The kernel, which occurs in the kinetic equation for the quantum-dot spin valve, is the sum of the contributions from the two leads, ${\bf W}=\sum_r {\bf W}_r$.
	Arranged in the basis ${\bf p}=(P_{0},P_{1},P_d,S_{x},S_{y},S_{z})$ the part ${\bf W}_L$ is given by $\GL$ times
	\begin{widetext}
		\begin{small}
			\begin{displaymath}
				\left(
				\begin{array}{cccccc}
					-2f_{L}(\varepsilon) & X f_{L}^-(\varepsilon) & 0 & 2 p_{L} X f_{L}^-(\varepsilon)\cos\frac{\phi}{2} &2 p_{L} X f_{L}^-(\varepsilon) \sin\frac{\phi}{2} & 0 \\
					2 X^{-1} f_{L}(\varepsilon) & -A^+ & 2Xf_{L}^-(\varepsilon+U)&  -2A^-p_{L} \cos\frac{\phi}{2} &   -2 A^-p_{L} \sin\frac{\phi}{2}&0 \\ 
					0 & X^{-1} f_{L}(\varepsilon+U)  & -2 f_{L}^-(\varepsilon+U) &- 2 p_{L}X^{-1} f_{L}(\varepsilon+U)\cos\frac{\phi}{2}  & -2 p_{L}X^{-1} f_{L}(\varepsilon+U)\sin\frac{\phi}{2} & 0 \\
					 p_{L}X^{-1} f_{L}(\varepsilon)\cos\frac{\phi}{2} & -\frac{p_{L}}{2} A^-\cos\frac{\phi}{2} &- p_{L} X f_{L}^-(\varepsilon+U)\cos\frac{\phi}{2}  & -A^+ & 0 & p_{L}\beta_{L}\sin\frac{\phi}{2}  \\
					 p_{L}X^{-1} f_{L}(\varepsilon)\sin\frac{\phi}{2} &  -\frac{p_{L}}{2} A^-\sin\frac{\phi}{2}  & -p_{L} X f_{L}^-(\varepsilon+U)\sin\frac{\phi}{2}  & 0 & -A^+ &p_{L}\beta_{L}\cos\frac{\phi}{2}  \\
					0  & 0 & 0 & p_{L}\beta_{L}\sin\frac{\phi}{2}  &p_{L}\beta_{L}\cos\frac{\phi}{2}  & -A^+
				\end{array}
				\right).
			\end{displaymath}
		\end{small}
	\end{widetext}
	For a clearer presentation we have made the following definitions: $f_{L}^-(\varepsilon)\equiv1-f_{L}(\varepsilon)$, $A^\pm \equiv f_{L}^-(\varepsilon)\pm f_{L}(\varepsilon+U)$, $X\equiv \text{exp}(i\chi)$, and $\beta_{L}\equiv |{\bf B}_{L}|/(p_{L}\GL)$, with Fermi function $f(E)=[\exp(E/k_BT)+1]^{-1}$ and $|{\bf B}_{L}|$ being the absolute value of the exchange field of the left lead, see Ref.~\onlinecite{braun:2004}. To obtain the part ${\bf W}_R$ the replacements $L\rightarrow R$ and $\phi\rightarrow-\phi$ have to be made.
	
	For the calculations concerning the F-D-N system we used the kernel ${\bf W}_{\text{FDN}}(\chi_\up,\chi_\dn)$, which allows for spin-resolved counting. In the basis ${\bf p}=(P_{0},P_{\up},P_{\dn},P_d)$ it is
	\begin{widetext}
		\begin{small}
			\begin{displaymath}
				{\bf W}_{\text{FDN}}(\chi_\up,\chi_\dn)=\sum_r\left(
				\begin{array}{cccc}
					-\Gr f_{r}(\varepsilon) & \Gamma_{r,\up}X_\up f^-_{r}(\varepsilon) & \Gamma_{r,\dn}X_\dn f^-_{r}(\varepsilon) & 0 \\
					\Gamma_{r,\up}X^{-1}_\up f_{r}(\varepsilon) & -\left[\Gamma_{r,\dn}f_r(\varepsilon+U)+\Gamma_{r,\up}f_r^-(\varepsilon)\right] & 0 & \Gamma_{r,\dn}X_\dn f_{r}^-(\varepsilon+U) \\ 
					\Gamma_{r,\dn}X^{-1}_\dn f_{r}(\varepsilon) & 0 &-\left[\Gamma_{r,\up}f_r(\varepsilon+U)+\Gamma_{r,\dn}f_r^-(\varepsilon)\right] & \Gamma_{r,\up}X_\up f_{r}^-(\varepsilon+U)  \\
					0 &  \Gamma_{r,\dn}X^{-1}_\dn f_{r}(\varepsilon+U)  & \Gamma_{r,\up}X^{-1}_\up f_{r}(\varepsilon+U) & -\Gr f^-_r(\varepsilon+U)
				\end{array}
				\right),
			\end{displaymath}
		\end{small}
	\end{widetext}
	with $X_{\up,\dn}\equiv \text{exp}(i\chi_{\up,\dn})$.

\end{document}